\documentclass[aps,prl,floatfix,twocolumn,showpacs,superscriptaddress,groupedaddress]{revtex4-2}  
\usepackage{graphicx}  
\usepackage{amsmath,amssymb}
\usepackage{siunitx}
\usepackage{flushend}
\usepackage{array}
\usepackage{float}
\usepackage{natbib}
\usepackage{booktabs}
\usepackage{color}

\begin{document}

\title{Berry Curvature Dipole and its Strain Engineering in Layered Phosphorene}

\author{Arka Bandyopadhyay} \email[]{arkabandyopa@iisc.ac.in}
\affiliation{Solid State and Structural Chemistry Unit, Indian Institute of Science, Bangalore 560012, India.}
\author{Nesta Benno Joseph}
\affiliation{Solid State and Structural Chemistry Unit, Indian Institute of Science, Bangalore 560012, India.}
\author{Awadhesh Narayan} \email[]{awadhesh@iisc.ac.in} 
\affiliation{Solid State and Structural Chemistry Unit, Indian Institute of Science, Bangalore 560012, India.}

\vskip 0.25cm

\date{\today}

\vskip 0.25cm
\begin{abstract}
The emergence of the fascinating non-linear Hall effect intrinsically depends on the non-zero value of the Berry curvature dipole. In this work, we predict that suitable strain engineering in layered van der Waals material phosphorene can give rise to a significantly large Berry curvature dipole. Using symmetry design principles, and a combination of feasible strain and staggered on-site potentials, we show how a substantial Berry curvature dipole may be engineered at the Fermi level. We discover that monolayer phosphorene exhibits the most intense Berry curvature dipole peak near 11.8\% strain, which is also a critical point for the topological phase transition in pristine phosphorene. Furthermore, we have shown that the necessary strain value to achieve substantial Berry curvature dipole can be reduced by increasing the number of layers. We have revealed that strain in these van der Waals systems not only alters the magnitude of Berry curvature dipole to a significant value but allows control over its sign. We are hopeful that our predictions will pave way to realize the non-linear Hall effect in such elemental van der Waals systems.
\end{abstract}

\maketitle

\date{\today}

\section{Introduction}

The quantization of Hall conductance in a strong magnetic field~\cite{thouless1982quantized,niu1987quantum} and its robustness to external perturbations have introduced several intricate concepts of topology~\cite{klitzing1980new,qi2011topological,hasan2010colloquium} in condensed matter physics. It has been observed that breaking of the time-reversal symmetry essentially gives rise to non-trivial geometric properties of the electronic wavefunction~\cite{haldane2004berry,xiao2010berry}. This non-trivial nature of the Bloch states encoded in the Berry curvature, in turn, protects the plateaus in the quantum Hall effect (QHE)~\cite{cage2012quantum,chang2023colloquium}. In particular, each Hall plateau can be associated with a specific integer, i.e., the Chern number, which is the momentum-integrated Berry curvature over the Brillouin zone for the occupied bands~\cite{bernevig2013topological}. The Berry curvature transforms as an odd function of the crystal momentum under time-reversal symmetric conditions. Therefore, in this case, the integration over the Brillouin zone, mentioned above, will always be trivially zero resulting in the disappearance of the Hall conductance. However, this argument is valid, only when one considers the linear variation of the Hall voltage with the external electric field. Beyond the linear response regime, the positive and negative values of Berry curvature may segregate out to distinct momentum values -- causing a dipole-like term popularly known as the Berry curvature dipole (BCD)~\cite{sodemann2015quantum}. The BCD is intrinsically protected by the broken inversion symmetry and is the primary ingredient to evince a non-linear or second-order Hall effect under the influence of an external electric field~\cite{du2021nonlinear,ortix2021nonlinear}.

The crystal symmetry plays a crucial role in predicting the fate of such a nonlinear Hall response in any inversion-symmetry broken system. For instance, a single mirror line is the maximum symmetry allowed for a nonzero value of BCD in any two-dimensional (2D) crystal. In this regard, various low-symmetry systems that exhibit massive tilted Dirac cones~\cite{nandy2019symmetry,joseph2021tunable,du2018band,joseph2021topological} or Weyl cones~\cite{zeng2021nonlinear,kumar2021room,roy2022non,singh2020engineering,zhang2018berry,zhuang2022extrinsic} have been intensively investigated to achieve sizeable BCD. Recent exciting experiments also reveal that the BCD-induced non-linear Hall effect can not only be measured but also externally controlled in strained or layered transition-metal dichalcogenides, i.e., WTe$_2$~\cite{ma2019observation,kang2019nonlinear,xu2018electrically,xiao2020berry,ye2023control}, MoS$_2$~\cite{lee2017valley,son2019strain}, and WSe$_2$~\cite{huang2022giant}. Furthermore, it has been observed that the warping of the Fermi surface can also trigger appreciable BCD in graphene, which is comparable to that of transition-metal dichalcogenides~\cite{battilomo2019berry}. Furthermore, significantly large BCD has been experimentally observed in corrugated bilayer graphene~\cite{ho2021hall} and twisted double bilayer graphene, i.e., moir{\'e} superlattices~\cite{sinha2022berry}. Beyond graphene, an electrically switchable giant BCD has been predicted in buckled honeycomb lattices, namely silicene, germanene, and stanene~\cite{bandyopadhyay2022electrically}. It is worth mentioning that the search for the nonlinear Hall effect in material systems has also been extended to three-dimensional topological semimetals WTe$_{2}$ and Cd$_{3}$As$_{2}$ \cite{shvetsov2019nonlinear}. The exciting concepts of BCD laid the foundation of the intrinsic part of the unconventional second-order electronic and thermal Hall effect. Therefore, predicting new feasible materials with large BCD is an active field of research that could lead to fascinating practical applications in rectification~\cite{isobe2020high}, terahertz detectors~\cite{zhang2021terahertz}, and opto-spintronic devices~\cite{habara2022nonlinear}. 

\begin{figure*}[!hbt]
\centering
\includegraphics[scale=0.3]{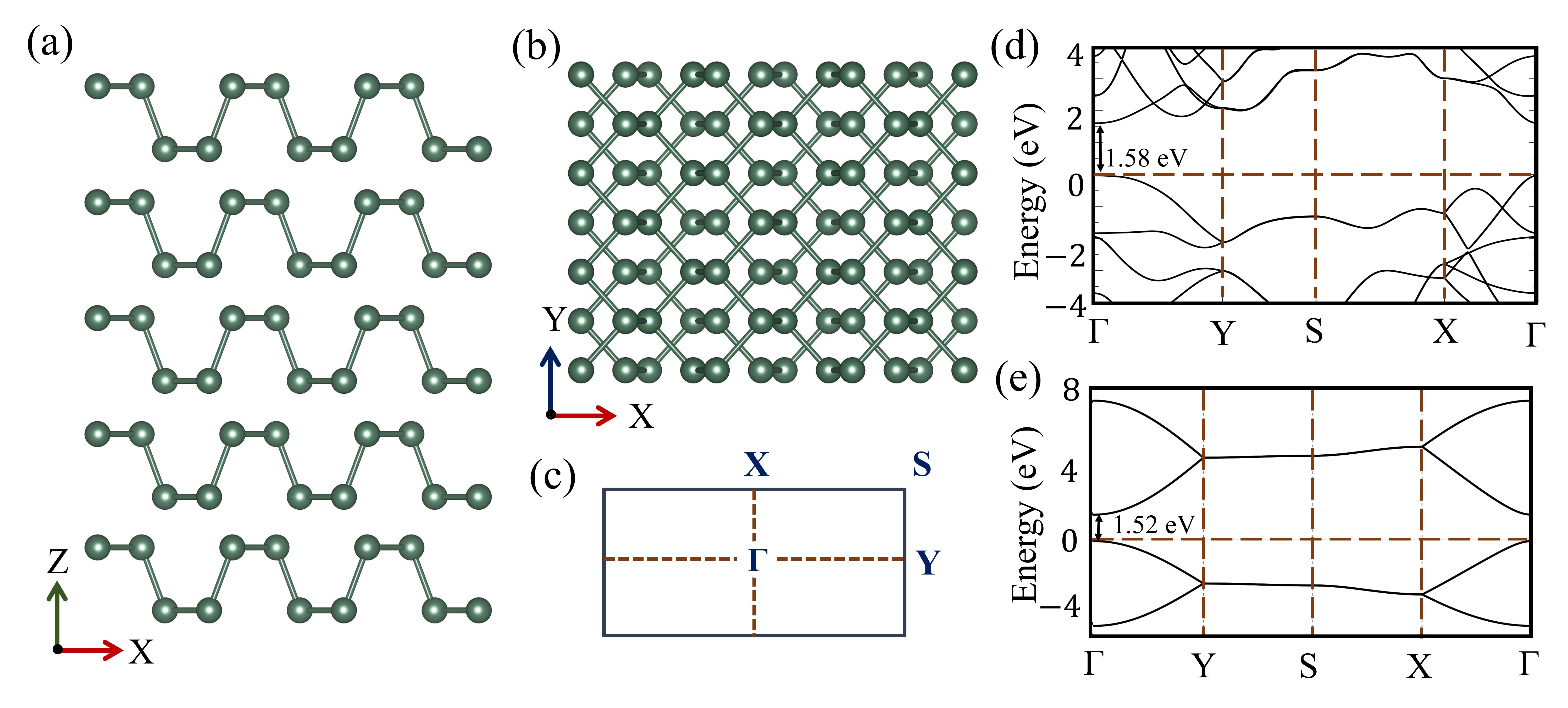}
\caption{\textbf{Crystal structure and electronic band structure of phosphorene.} (a) Side view and (b) top view of the layered phosphorene structure. (c) The Brillouin zone and the high symmetry points of phosphorene. (d) HSE band structure for monolayer phosphorene structure reveals a band gap of nearly 1.58 eV at the $\Gamma$ point. (e) The band structure obtained from our tight-binding model exhibits a band gap of 1.52 eV, that agrees well with the HSE calculation.} \label{fig:structure}
\end{figure*}

Similar to graphite, black phosphorus is a layered system in which adjacent layers are stacked together by van der Waals interactions. As a consequence, single- and multi-layer black phosphorus (phosphorene) was first successfully exfoliated using a scotch-tape microcleavage method in 2014~\cite{li2014black}. Subsequently, phosphorene has drawn considerable research attention owing to the layer-dependent tunable direct bandgap, significant carrier mobility, and in-plane anisotropy, among many other exciting properties~\cite{liu2015semiconducting,carvalho2016phosphorene}. In particular, the bandgap of bulk black phosphorus is $\approx$ 0.33 eV, which increases with the reduction of number of layers and finally reaches a value $\approx$ 2.0 eV for the single layer phosphorene~\cite{tran2014layer}. This tunable direct band gap of phosphorene overcomes the fundamental drawback of graphene, and, therefore, several efforts have been made to use phosphorene in multiple nanodevice applications, including transistors, optoelectronics, and batteries~\cite{chaudhary2022phosphorene,glavin2020emerging}. In experiments, phosphorene-based field effect transistors exhibit significantly large on-current, hole field-effect mobility, and on-off ratio even at room temperature~\cite{glavin2020emerging}. Further, a ballistic quantum transport providing a much faster switching speed can be attained by reducing the number of layers or making superstructures of phosphorene~\cite{cheng2016anisotropic,guzman2023disorder}. Moreover, the phosphorene structure exhibits the QHE that carries the fingerprint of a bias-driven topological phase transition~\cite{ma2017quantum,yuan2016quantum,liu2015switching,dolui2015quantum}. 

Even though phosphorene exhibits all of these fascinating properties, its inherent instability under ambient conditions restricts practical applications. To overcome this drawback, an encapsulation technique has been adopted where the phosphorene layers are passivated by various capping materials before exposing it to air. For example, Al$_2$O$_3$ layers not only limit the rapid oxidation but also enhance its electronic responses~\cite{galceran2017stabilizing}. Similarly, phosphorene can be safely sandwiched between stable graphene and hexagonal boron nitride sheets~\cite{doganov2015transport}. Furthermore, there exist a wide variety of material systems ranging from SnO$_{2}$, TiO$_{2}$, SiO$_2$ to layers to different polymers that suppress the degradation process and improve the duration and stability of phosphorene~\cite{li2019tunable,sang2019recent,kim2015toward,edmonds2015creating,yasaei2015stable}.

In this work, we predict that single and multilayered phosphorene are promising platforms to suitably engineer a large BCD at the Fermi level. For this purpose, we propose crystal symmetry reduction and inversion symmetry breaking, which can be readily achieved using a feasible staggered onsite potential. Furthermore, we introduce a strain to tune the band structure of phosphorene, particularly the band gap. We systematically examine the combined effect of this symmetry reduction and strain-generated band gap modulation on the BCD of the system. We demonstrate that the layered phosphorene reduces the strain required for achieving large BCD peaks at the Fermi level compared to the single-layer case. We have essentially shown a significantly large BCD, whose value as well as sign can be modulated, in single and multilayered phosphorene sheets with broken inversion symmetry and under external strain. Our results put forth an appropriately perturbed phosphorene system as a promising material platform for various transport applications beyond linear response.

\section{Computational details}

Our first-principles calculations for monolayer phosphorene were carried out based on the density functional theory (DFT) framework as implemented in the {\sc quantum espresso} code~\cite{QE-2017,QE-2009}. A kinetic energy cut-off of $30$ Ry was considered, using the ultrasoft pseudopotentials~\cite{PhysRevB.41.7892} to describe the core electrons. The monolayer was modelled with a vacuum region of $\sim$10 \AA ~along the $c$ axis to reduce any interaction between periodic images. 
The Brillouin zone was sampled over a uniform $\Gamma$-centered $k$-mesh of $9\times6\times1$. In order to obtain an accurate band gap, consistent with our model calculations, the Heyd–Scuseria–Ernzerhof (HSE)~\cite{heyd2003hybrid} screened hybrid functional was used with a q-mesh of $3\times2\times1$ for sampling the Fock operator. In case of phosphorene layers, we have used "Grimme-d2" type van der Waals correction in our DFT study~\cite{grimme2006semiempirical}. The choice of "Grimme-d2" type van der Waals correction to the very accurate HSE hybrid functional is well-equipped to explore the essential physics for generating significant BCD in the strain-engineered phosphorene layers. The reliability of our numerical method has been confirmed by comparing the first-principle band gap value with the experimental one~\cite{woomer2015phosphorene,li2017direct}. In this regard, it is worth mentioning that the in-plane biaxial strain in phosphorene effectively tunes the van der Waals interaction of the layered phosphorene structure that weakens the coupling between the adjacent layers. This signature has been experimentally captured using infrared spectroscopy~\cite{huang2019strain,li2017direct,zhang2017infrared,kim2021actively,akhtar2017recent}.

Our numerical calculations of the Berry curvature and the BCD have been efficiently performed using the Wannier-Berri package~\cite{tsirkin2021high} and the PythTB module~\cite{Pythtb}. The "Python Tight Binding" or "PythTB" code essentially solves the tight-binding models for the electronic structure. In particular, the PythTB module computes electron eigenvalues and eigenvectors for a selected $k$ mesh in the Brillouin zone. This can further be applied to evaluate the topological properties such as the Berry phase, Berry connection, and Berry curvature. On the other hand, the Wannier-Berri code is a highly efficient Wannier interpolation tool that can evaluate the $k$-space integrals of Berry curvature, orbital moment, and other derived quantities employing maximally localized Wannier functions or tight-binding models. Therefore, Wannier-Berri is a handy tool to evaluate the Berry curvature dipole, which is the key focus of our work. We first write down the realistic tight-binding model in momentum space for our layered phosphorene system in the presence of external strain and inversion symmetry breaking potential. After that the eigenvalues of the tight binding Hamiltonian are evaluated using the PythTB module that provides the electronic band structure of the system. The tight binding Hamiltonian written in the PythTB module serves as an input for the Wannier-Berri code. Wannier-Berri computes Berry curvature dipole in two ways, i.e., the Fermi surface integral and the Fermi sea integral, which are given as

\begin{equation}
  D_{ab}^{surf} = - \int_{BZ} d\vec{k} \sum_{n} \left[ \frac{\partial}{\partial k^{a}} f_0(\epsilon_{nk}) \right] \Omega_{nk}^{b},
\end{equation}

and

\begin{equation}
  D_{ab}^{sea} = \int_{BZ} d\vec{k} \sum_{n} f_0(\epsilon_{nk}) \left( \frac{\partial}{\partial k^{a}}  \Omega_{nk}^{b} \right).
\end{equation}

As expected, these approaches yield the same result.

\begin{figure*}[!hbt]
\centering
\includegraphics[scale=0.25]{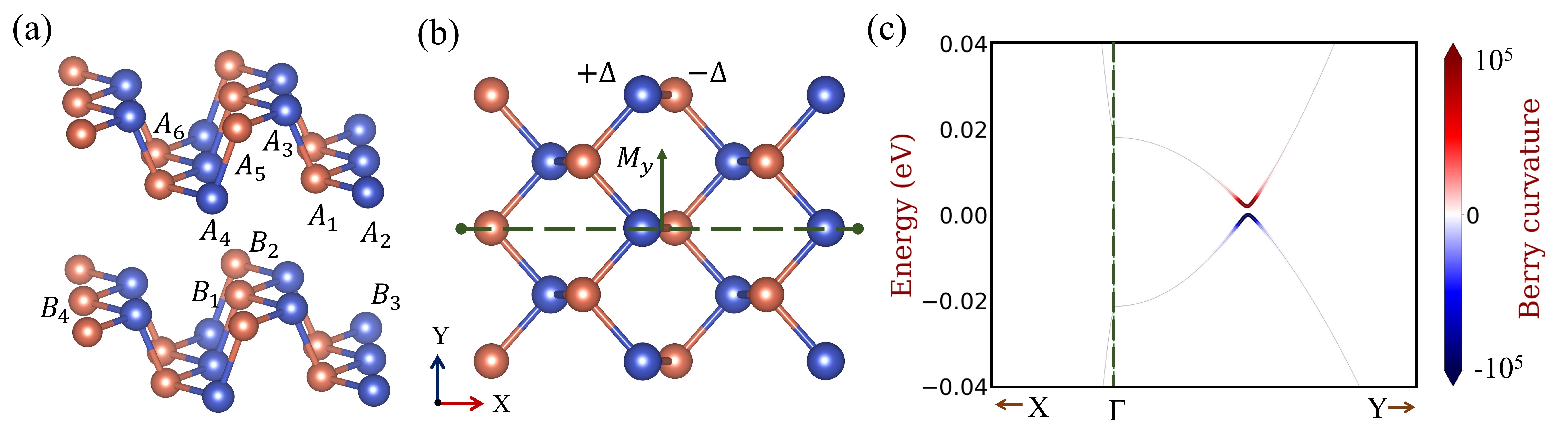}
\caption{\textbf{Inversion symmetry breaking and Berry curvature in strained phosphorene.}(a) The tight binding lattice of layered phosphorene. Here two layers are shown for ease of visualization. The intralayer hopping parameters between $A_{1} \leftrightarrow A_{2}$, $A_{1} \leftrightarrow A_{3}$, $A_{1} \leftrightarrow A_{4}$, $A_{1} \leftrightarrow A_{5}$, and $A_{2} \leftrightarrow A_{5}$ are $t_{10}$, $t_{20}$, $t_{30}$, $t_{40}$, and $t_{50}$, respectively. The interlayer hopping parameters between $A_{4} \leftrightarrow B_{1}$, $A_{6} \leftrightarrow B_{2}$, $A_{6} \leftrightarrow B_{4}$, and $A_{4} \leftrightarrow B_{4}$ are $g_{10}$, $g_{20}$, $g_{30}$, and $g_{40}$, respectively. (b) Inversion symmetry breaking can be introduced by considering a small energy offset ($2 \Delta$) between the red and blue atoms. The only preserved symmetry, in this case, is a single mirror line or a mirror plane $M_y$ that is orthogonal to the system plane. (c) The applied strain reduces the band gap of monolayer phosphorene, which gives rise to strong Berry curvature at the gapped Dirac-like conduction and valence bands. The value of applied strain for this band structure is 11.8\%. The positive and negative Berry curvatures are plotted with red and blue colour within the range $\pm 10^{5}$. Note that the band gap and the Berry curvature are tunable by the strain.} \label{fig:inversionbreaking}
\end{figure*}

\section{Tight-binding Model and symmetry analysis}

Graphene can be exfoliated from the three-dimensional hexagonal structure graphite, which has a P63/mmc symmetry with an associated point group of 6/mmm. The point group can be understood from the presence of one six-fold rotational symmetry and three mirror symmetry operations. Because of the planar honeycomb lattice structure, graphene possesses the highest symmetry D$_{6h}$ corresponding to one six-fold in-plane rotation, six two-fold perpendicular axes, and one horizontal mirror plane~\cite{battilomo2019berry}. The rotational symmetry of graphene can be primarily reduced in two possible ways. First, one can incorporate buckling, i.e., shifting one sublattice in the out-of-plane direction compared to the other in the system similar to silicene, germanene, stanene, arsenene, and antimonene~\cite{grazianetti2021rise}. These buckled honeycomb lattices exhibit one three-fold rotation, three two-fold perpendicular axes, and three mirror planes generating the $D_{3d}$ point group~\cite{bandyopadhyay2022electrically}. As a consequence, the in-plane properties of these systems are isotropic. Alternatively, considering a puckered lattice structure such as black phosphorus also reduces the crystal symmetry. Monolayer phosphorene with symmetry to D$_{2h}$ exhibits a rectangular unit cell that consists of four P atoms. Different numbers of phosphorene layers can be stacked over each other, as depicted in Fig.~\ref{fig:structure} (a) and (b). In this regard it it worth mentioning that, we have considered the two-dimensional phosphorene with symmetry group D$_{2h}$ constructed from the bulk phosphorene structure with Cmce space group in our DFT study. It is evident that the $C_3$ symmetry is inherently lost, so the remaining symmetry elements are one two-fold rotation, one mirror plane, and two two-fold perpendicular axes. Unlike graphene and D$_{3d}$ systems, the black phosphorous structure demonstrates highly anisotropic optical, electronic, and transport properties. We know that the emergence of the nonlinear Hall effect in two dimensions is permitted for a maximum of a single mirror plane. Therefore, black phosphorus or phosphorene is a natural choice to evince a non-zero Berry curvature dipole induced nonlinear Hall effect because of a low crystal symmetry. The phosphorene system is also invariant under spatial inversion operation. Therefore, we have reduced the crystal symmetry of phosphorene by introducing an inversion symmetry breaking potential, which may arise as the effect of a substrate.
For our calculations, we introduce a tight-binding model for the layered phosphorene, as given below

\begin{equation}
    H = \sum_{i} \epsilon_i \: c_{i}^{\dagger} c_{i} + \sum_{i \neq j} t_{ij} \: c_{i}^{\dagger} c_{j} + \sum_{i \neq j} g_{ij} \: c_{i}^{\dagger} c_{j}.
    \label{eq:tightbinding}
\end{equation}

Here $\epsilon_i$, $t_{ij}$, and $g_{ij}$ represent the onsite energy, the in-plane and the out-of-plane hopping parameters, respectively. In the case of pristine monolayer phosphorene, the onsite energy $\epsilon_i$ for all P atoms in the unit cell can be considered to be uniform, and the interlayer hopping parameters $g_{ij} = 0$. The in-plane hopping parameters can be determined by reproducing the electronic band structure, particularly the band gap along the high symmetry path of the Brillouin zone shown in Fig.~\ref{fig:structure} (c). For this purpose, we have calculated the band structure using the HSE hybrid functional that exhibits a direct band gap of 1.58 eV at $\Gamma$ point [Fig.~\ref{fig:structure} (d)]. It is worth mentioning that this band gap is in close agreement with the previous report of 1.51 eV by Qiao \textit{et al.}~\cite{qiao2014high}. Our tight-binding Hamiltonian is similar to the one proposed by Rudenko \textit{et al.}~\cite{rudenko2014quasiparticle}, which successfully reproduces the band gap (1.52 eV) for the hopping parameters $t_{10} = -1.220$ eV, $t_{20} = 3.665$ eV, $t_{30} = -0.205$ eV, $t_{40} = -0.105$ eV, and $t_{50} = -0.055$ eV, as given in Fig.~\ref{fig:structure} (e). The intralayer and interlayer hopping parameters for monolayer and multilayer phosphorene are schematically presented in Fig.~\ref{fig:inversionbreaking} (a). It is clear from Fig.~\ref{fig:inversionbreaking} (a) that the intralayer hopping parameters between $A_{1} \leftrightarrow A_{2}$, $A_{1} \leftrightarrow A_{3}$, $A_{1} \leftrightarrow A_{4}$, $A_{1} \leftrightarrow A_{5}$, and $A_{2} \leftrightarrow A_{5}$ are $t_{10}$, $t_{20}$, $t_{30}$, $t_{40}$ and $t_{50}$, respectively. Further, for multilayered phosphorene, the interlayer hopping parameters are given by $A_{4} \leftrightarrow B_{1}$, $A_{6} \leftrightarrow B_{2}$, $A_{6} \leftrightarrow B_{4}$, and $A_{4} \leftrightarrow B_{4}$, which are $g_{10}$, $g_{20}$, $g_{30}$, and $g_{40}$, respectively. The values for interlayer hopping parameters $g_{10}$, $g_{20}$, and $g_{30}$ can also be similarly evaluated, which are 0.295 eV, 0.273 eV, $-0.091$ eV, and $-0.151$ eV, respectively. Needless to say, the presence of inversion symmetry in the phosphorene forbids the system from exhibiting a non-zero Berry curvature. For that purpose, we have intuitively assigned different onsite energies ($\pm \Delta$) to the blue and red atoms, as shown in Fig.~\ref{fig:inversionbreaking} (b). Such an inversion symmetry breaking term can occur in the Hamiltonian due to the combined out-of-plane and in-plane lattice commensurate electric fields or the presence of monochalcogenide substrate~\cite{hu2019recent,low2015topological,yar2023nonlinear}.
To establish this, we have performed the DFT computations for the phosphorene system with the SnSe substrate, as depicted in Fig. 3. In recent work, Muzaffar \textit{et al.}~\cite{muzaffar2021epitaxial} have explored that phosphorene is energetically and thermodynamically stable on the SnSe(001) substrate. The corresponding lattice mismatches along the armchair and zigzag directions are 1.32\% and 1.07\%, respectively, which are low. We have found that there is no inversion symmetry in the system. The only symmetry that survives even after the presence of substrate generated potential is a single mirror plane $M_y$, thus enabling a non-zero BCD.
\begin{figure}[t]
    \centering
    \includegraphics[width=8cm]{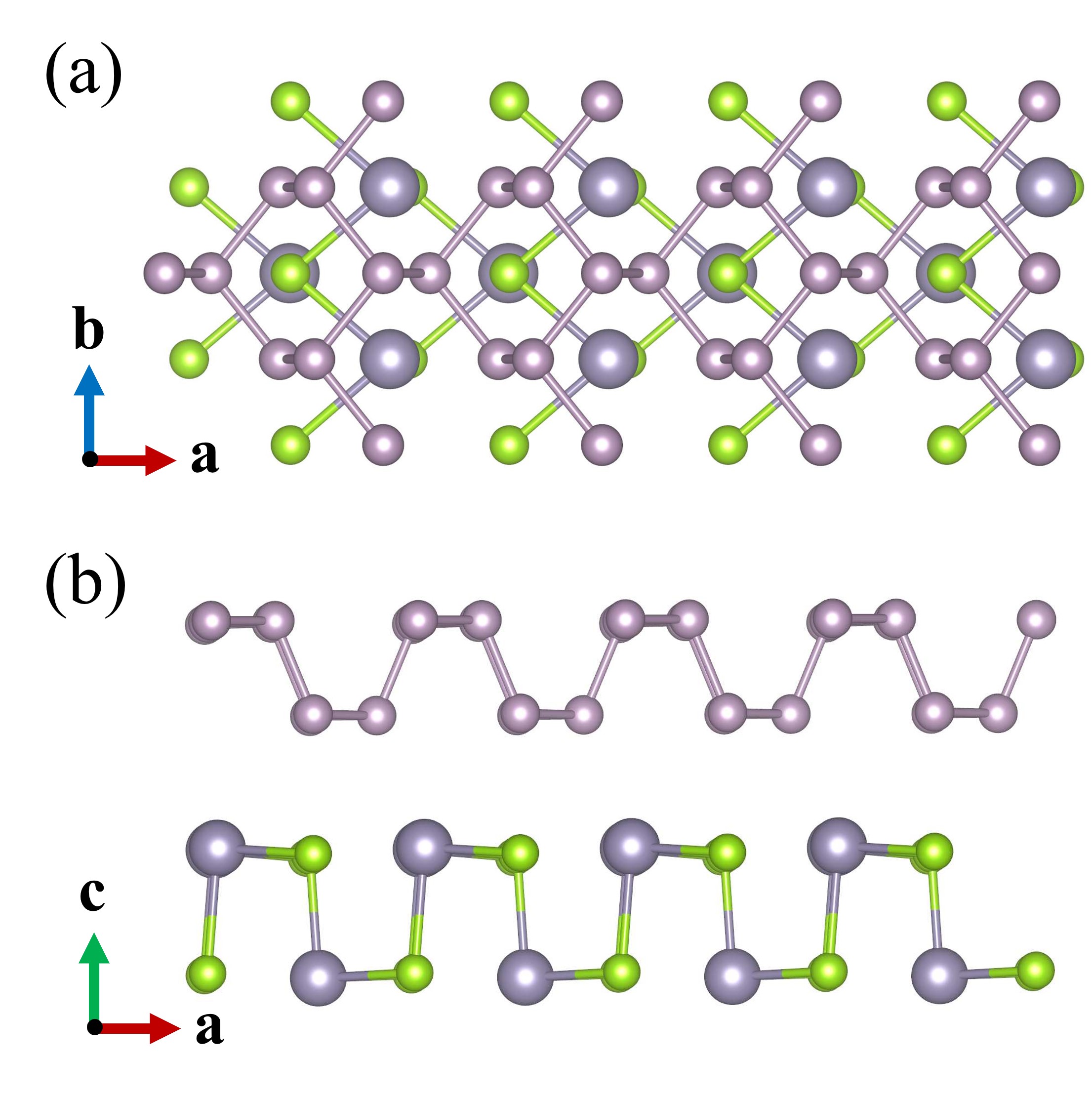}
    \caption{\textbf{Phosphorene on the monochalchogenide SnSe substrate.} (a) Top and (b) side view of phosphorene on substrate monochalcogenide SnSe. The phosphorene experiences an inversion symmetry breaking potential as considered in the tight-binding model. The crystal symmetry is also reduced to a single mirror plane $M_y$.}
    \label{fig:substrate}
\end{figure}
Incorporating such an asymmetric onsite energy reduces the symmetry of the system down to a single mirror line $M_y$. We note that the ocurrence of BCD in such a system is no longer restricted by the crystallographic symmetry. However, the wide band gap of inversion symmetry broken phosphorene does not allow any BCD response near the Fermi level. We next address this issue by proposing a strain engineering approach.

\begin{table}[!hbt]
    \centering
     \caption{The $\alpha$ coefficients for different hopping strengths in phosphorene obtained from the structure parameters.}
    \setlength{\tabcolsep}{10pt} 
\renewcommand{\arraystretch}{1.5} 
     \begin{tabular}{c|c c c}
     \hline
     \hline
hopping & $\alpha_x$ & $\alpha_y$ & $\alpha_z$ \\
\hline
\hline
$t_{10}$ & 0.4460 & 0.5571 & 0.0000 \\
$t_{20}$ & 0.0992 & 0.0000 & 0.9052 \\
$t_{30}$ & 0.7505 & 0.2461 & 0.0000 \\
$t_{40}$ & 0.3976 & 0.2280 & 0.3722 \\
$t_{50}$ & 0.7530 & 0.0000 & 0.2538 \\
\hline
\hline
    \end{tabular}
      \label{tab:coefficient}
\end{table}

\section{BCD in strained monolayer phosphorene}

The above discussions show that band gap reduction is essential to induce appreciable BCD near the Fermi level in phosphorene. One of the efficient ways to reduce the band gap in phosphorene is strain engineering. Phosphorene is well-known to exhibit superior mechanical flexibility even up to 30\% strain~\cite{wei2014superior,peng2014strain,wang2015strain}. Considering the linear deformation region and the Harrison rule~\cite{jiang2015analytic,sisakht2016strain} for the dependency of the $p$ orbital hopping parameter on the bond length, we can write the following expression for the modified hopping parameter

\begin{equation}
    t_i \approx [1 - 2 (\alpha_{x}^{i} \epsilon_x + \alpha_{y}^{i} \epsilon_y+ \alpha_{z}^{i} \epsilon_z)] t_{i0}.
\label{eq:modifiedhopp}
\end{equation}

Here $t_i$ and $t_{i0}$ represent the modified and original hoppping parameters, respectively, $\epsilon_{x/y/z}$ is the strain along $x/y/z$ direction. The coefficient $\alpha^{n}_{m}$ can be evaluated from the structure parameters~\cite{zhao2016topological,yarmohammadi2020anisotropic}, which are tabulated in Table~\ref{tab:coefficient}. 

It is worth mentioning that these intralayer and interlayer hopping parameters accurately reproduce the first-principles band gap of strained phosphorene in the presence of van der Waals interactions. For example, the band gap values at different strains in the monolayer phosphorene system calculated using both DFT and tight binding results are listed in Table~\ref{table}.

\begin{table}[H]
\setlength{\tabcolsep}{14pt}
\centering
\caption{Variation of band gap of monolayer phosphorene with strain along crystallographic a-axis evaluated from hybrid DFT functional HSE and tight-binding (TB) model.}\label{table}
\begin{tabular}{c|cc}
\hline
\hline
strain                  & HSE band gap                    & TB band gap  \\ 
(\%)                    & (eV)                  &  (eV) \\
\hline
\hline
0                     & 1.6                     & 1.5  \\
2                     & 1.7                     & 1.6   \\
4                     & 1.8                     & 1.7  \\
6                     & 1.9                     & 1.8  \\
8                     & 2.0                     & 1.9  \\
\hline
\hline
\end{tabular}
\end{table}

The variation of Berry curvature and its first moment (BCD) strongly depends on the band gap tuning of the system with an external strain. Moreover, we have found that the band gap decreases with increasing tensile strain $\epsilon_z$ in pristine phosphorene ($\Delta = 0$). The band gap finally vanishes for a critical strain $\epsilon_{zc} = 11.8\%$. Further increase in strain value causes a band inversion in the system. In other words, the tensile strain $\epsilon_z$ drives the monolayer phosphorene system through a topological phase transition around the critical value $\epsilon_{zc}$~\cite{sisakht2016strain}.

\begin{figure}[t]
\centering
\includegraphics[scale=0.28]{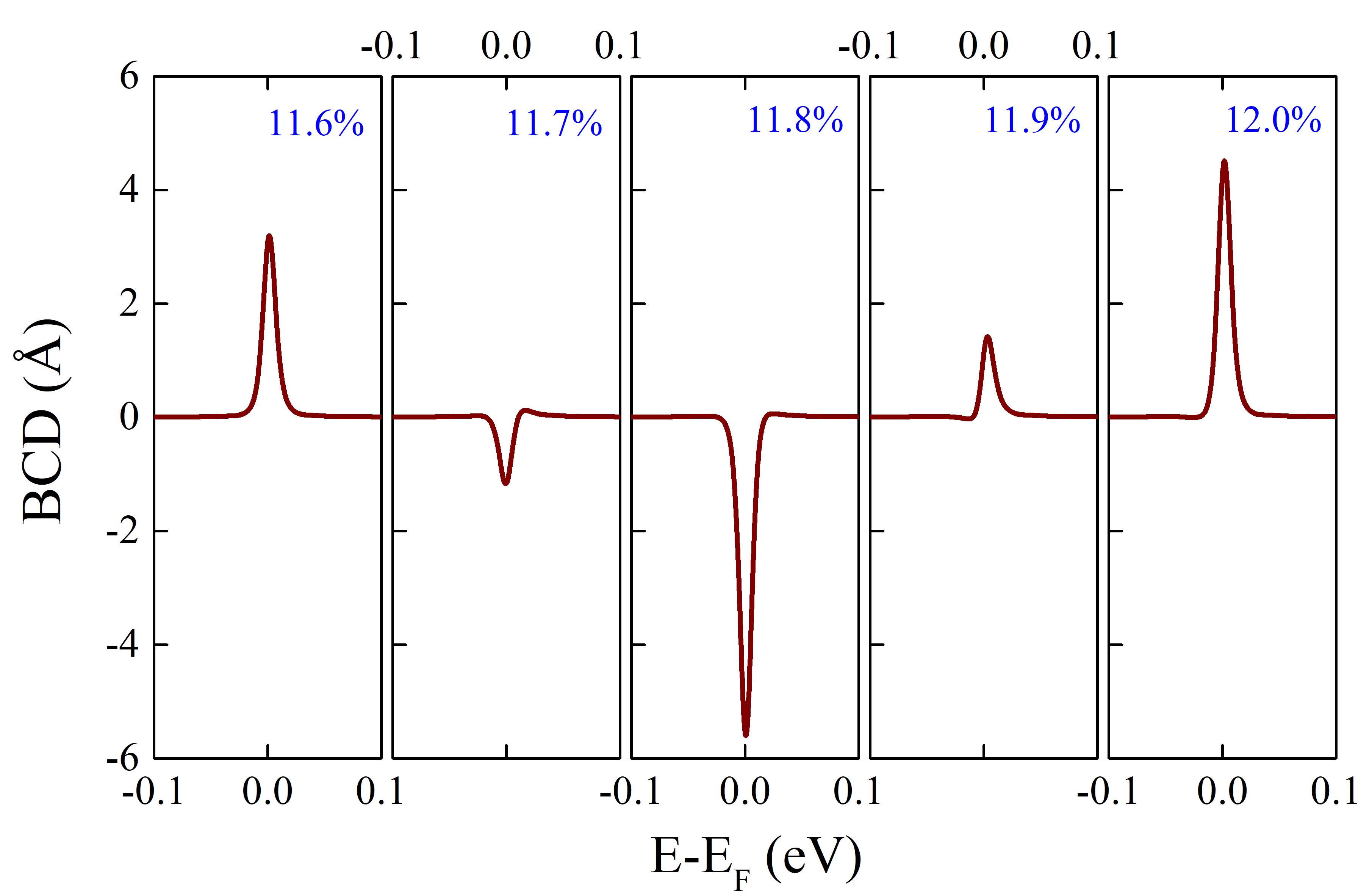}
\caption{\textbf{Variation of BCD with strain for monolayer phosphorene.} Inversion symmetry breaking induces two BCD peaks at the top of the valence band and bottom of the conduction band. With increasing strain the peaks move towards each other because of band gap reduction. Finally, the peaks merge and the monolayer phosphorene exhibits a large of BCD ($D_{yz}$ component) peak at the Fermi level for the external strain of 11.6\%. A change in the magnitude and sign of the BCD is observed beyond this strain value. The strain values corresponding to the BCD are shown in the plots.} \label{fig:bcd1}
\end{figure}

We have employed a similar approach in the case of phosphorene with a weak staggered onsite term $\Delta=0.001$ eV. We have found that such an inversion symmetry breaking allows the Berry curvature to be nonzero. Furthermore, the strength of the Berry curvature increases with increasing tensile strain $\epsilon_z$, as it reduces the band gap. It is worth mentioning that near the critical strain, $\epsilon_{zc} = 11.8\%$ discussed above, the Berry curvature is found to be the most pronounced because of the narrow band gap, as shown in Fig.~\ref{fig:inversionbreaking}(c). We note that the Berry curvature in a 2D crystal is only directed along the out-of-plane direction (the $z$ direction in our case). The Berry curvature value can be estimated using the Kubo formalism~\cite{xiao2010berry}. The Berry curvature given as

\begin{equation}
    \Omega_z (\vec{k}) = 2 i \sum_{i \neq j} \frac{\langle {i} |{\partial\hat{H}/\partial k_x} |{j} \rangle \langle{j}| {\partial\hat{H}/\partial k_y} |{i}\rangle}{\left(\lambda_i - \lambda_j\right)^{2}},
\label{eq:bc}
\end{equation}

where $\lambda_i$ and $\lambda_j$ are the eigenvalues of the tight-binding Hamiltonian $H$ and the corresponding eigenvectors are given by $|{i}\rangle$ and $|{j}\rangle$, respectively. Moreover, the differentiation of the Berry curvature (of the occupied states) with respect to the momentum will contribute to the nonlinear conductivity tensor ($\chi_{\alpha \beta \gamma}$) as

\begin{equation}
    \chi_{\alpha \beta \gamma} = - \epsilon_{\alpha \delta \gamma} \frac{e^3 \tau}{2(1+i\omega \tau)} \int_{k} [d\vec{k}] f_0 \left(\frac{\partial \Omega_\delta}{\partial k_\beta}\right),
    \label{eq:chi}
\end{equation}

where $e$, $\tau$, $\epsilon_{\alpha \delta \gamma}$, $f_0$ and $\Omega_\delta$ are the charge of an electron, scattering time, Levi-Civita symbol, equilibrium Fermi-Dirac distribution function, and $\delta$ component of the Berry curvature, respectively. The parameters $\alpha, \beta, \gamma, \delta \in \{x,y,z\}$ and in $d$ dimensions the $[d\vec{k}] = d^{d} k / (2 \pi)^{d}$. From Eq.~\ref{eq:chi}, the expression for the BCD tensor can be obtained as 

\begin{equation}
D_{\alpha \beta} = \int_{k} [d\vec{k}] f_0 \left(\partial \Omega_\beta / \partial k_\alpha \right).
\end{equation}

In other words, this BCD term triggers a nonlinear Hall current $J_{\alpha} = Re\{J_{\alpha}^{(0)} + J_{\alpha}^{(2\omega)} e^{i 2 \omega t}$\} under the influence of an oscillating electric field $\vec{E}(t) = Re\{\vec{E}_0 e^{i \omega t}\}$~\cite{sodemann2015quantum}. The current consists of both the rectified current ($J_{\alpha}^{(0)} = \chi_{\alpha \beta \gamma} \vec{E}_{\beta} \vec{E}_{\gamma}^{*}$) and a second harmonic term ($J_{\alpha}^{(2 \omega)} = \chi_{\alpha \beta \gamma} \vec{E}_{\beta} \vec{E}_{\gamma}$) with a doubled frequency $2 \omega$.

\begin{figure*}[!hbt]
\centering
\includegraphics[scale=0.17]{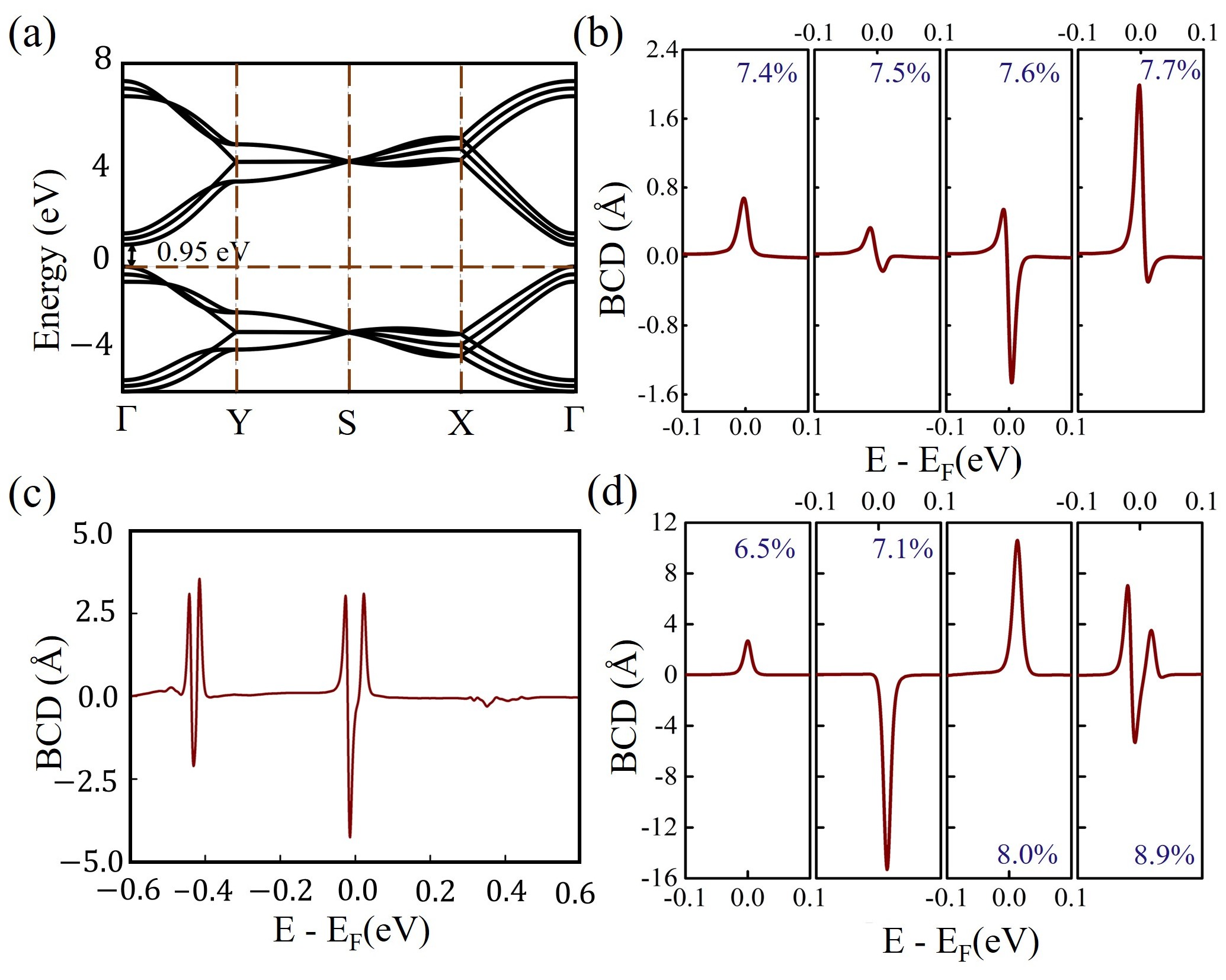}
\caption{\textbf{Effect of strain on the BCD in three- and five-layer phosphorene.} (a) Band structure of three-layer pristine phosphorene exhibits a band gap of 0.95 eV, which is significantly smaller than the monolayer. (b) Similar to the monolayer case, a sizable BCD ($D_{yz}$ component) is found at the Fermi level, which also gradually changes its sign upon introducing an applied tensile strain $\epsilon_z$. (c) Several large BCD peaks emerge near the Fermi level for the three-layer phosphorene at a strain value of $\approx$ 12\% owing to the formation of two Dirac-like cones at this strain value. (d) BCD of five-layer phosphorene exhibits several peaks at the Fermi level starting from the strain value of 6.5\%, which is smaller than that for the three-layer system. Similar to the other two cases, the BCD of five-layer phosphorene also changes its sign with increasing strain value. Finally, multiple BCD peaks appear near $\approx 8.9\%$ strain because of multiple Dirac-like band crossing points near the Fermi level.} \label{fig:bcd2}
\end{figure*}

After achieving such a large Berry curvature, we proceed to calculate the strain engineering effect on the BCD profile in monolayer phosphorene. The Berry curvature $\Omega_z (\vec{k})$ transforms as an odd function under time-reversal symmetry operation ($\mathcal{T}$) as $\mathcal{T}^{\dagger} \Omega_z(-\vec{k}) \mathcal{T} = - \Omega_z(\vec{k})$. Further, in the presence of the single mirror line $M_y$, $k_x$ and $k_y$ become even and odd functions, respectively. Therefore, it is evident that only the $d_{yz} = \partial \Omega_z / \partial y $ component will be an even function. Therefore, Eq.~\ref{eq:chi} reveals that the momentum integration of the above component $d_{yz}$ will provide the only nonzero component ($D_{yz}$) of the BCD. We discover that the value of the $D_{yz}$ component of the conduction and valence bands lying near the Fermi level increases with the band gap reduction. However, the BCD at the Fermi level is still absent because of the large band gap of the system. Finally, when the tensile strain $\epsilon_z$ approximately reaches a value of 11.6\%, these two independent BCD peaks merge, resulting in a substantially larger peak ($\approx 3.19$ \AA{}) at the Fermi level, as presented in Fig.~\ref{fig:bcd1}. As the strain is increased further, we find that the magnitude of BCD reaches a maximum value of $\approx 5.60$ \AA{} at the strain value of 11.8\%. Interestingly, the BCD peak sign also reverses compared to the value at 11.6\% strain. As we noted earlier, this 11.8\% tensile strain serves as a critical strain for the topological phase transition in pristine phosphorene. However, unlike a topological transition, in this inversion symmetry-breaking case, the BCD does not immediately switch around the 11.8\% stain, but a gradual change is observed, as we can see for 11.7\% and 11.9\% strain values in Fig.~\ref{fig:bcd1}. This difference between the BCD responses of topological transition and our case can be explained as follows -- the band structure of the inversion symmetry broken phase possesses a finite band gap even near 11.8\% strain because of the staggered onsite energy. Therefore, it does not allow any topological phase transition connected by a band gap closing point. We note that the primary effect of the topological phase transition would be to trigger a sudden switching of the Berry curvature dipole direction owing to the orbital exchange~\cite{bandyopadhyay2022electrically}.

\section{BCD in strained multilayer phosphorene}

In the above discussion, we have explored the strain engineering in monolayer phosphorene without an inversion center which gives rise to a significantly large BCD. Furthermore, the value and sign of the BCD at the Fermi level can also be tuned depending on the particular choice of the strain value. Next, we systematically analyze the effect of external strain on the BCD of multilayered phosphorene, particularly the representative cases of three and five layers. The primary advantage of the layered phosphorene over the monolayer is the relatively small band gap in the pristine case. Moreover, the number of bands with energy values close to the Fermi level also increases with increasing layer number. For example, the three-layer phosphorene consists of 12 P atoms in its unit cell and exhibits a band gap of 0.95 eV as shown in Fig.~\ref{fig:bcd2} (a). We find that the band gap decreases with increasing strength of tensile strain $\epsilon_z$. Furthermore, the inversion symmetry breaking using the staggered onsite potential allows a finite Berry curvature in the bands even under time-reversal symmetric conditions. In addition, the highly anisotropic band structure leads to a highly asymmetric Berry curvature distribution that, in turn, generates two non-zero BCD peaks near the top of the valence band and the bottom of the conduction band. Note that this occurrence of the BCD in three-layer phosphorene is allowed by the crystallographic symmetry of the system. These two weak BCD peaks move closer to each other and get strengthened with the introduction of strain. Finally, the two peaks merge to create a prominent peak at the Fermi level for the strain value of $ \epsilon_z \approx$ 7.4\% [Fig.~\ref{fig:bcd2} (b)]. As expected, the strain required for generating a large peak at the Fermi level is notably smaller in three-layer phosphorene as compared to the monolayer case (11.6\%). Further, we have observed a change in the sign of the BCD peak similar to the monolayer case between the strain values 7.4\% to 7.7\%. We have further increased the strain and found that the nature of the BCD curves at the Fermi level remains nearly unaltered, with the positive and negative BCD peaks (value $\approx \pm$ 4 \AA{}) at the valence and conduction bands adjacent to each other. However, several other smaller BCD peaks are also found away from the Fermi level because of the involvement of the remaining orbitals of the unit cell. Finally, the external strain modifies the band structure so that multiple Dirac-like bands occur near the Fermi level. These Dirac-like bands, by nature, result in multiple large BCD peaks near the Fermi level. It is clear from Fig.~\ref{fig:bcd2} (c) that 12\% strain in the three-layer phosphorene is sufficient to showcase such an exciting phenomenon of closely spaced large BCD peaks. In addition, some degenerate bands in the three-layer pristine phosphorene far from the Fermi level get separated because of inversion symmetry breaking. A considerable accumulation of Berry curvature is observed at these energy regions, which results in a strong BCD response. We demonstrate that the strain shifts these formerly degenerate energy values towards the Fermi level. Therefore, in addition to the Fermi level, another prominent BCD peak in present in the experimentally accessible range, i.e., 0.45 eV below the Fermi level for the strain value of $\approx$12\%. In the conduction band, i.e., $\approx$ 0.35 eV above the Fermi level, other noticeable BCD peaks occur because different bands in the conduction band approach each other near the above-mentioned strain values. 

All our findings discussed above for the three-layer case are even more pronounced in the five-layer system. The unit cell of the five-layer phosphorene contains 20 P atoms, so more bands are involved in producing the energy values near the Fermi level than the three-level system. As a result, the direct band gap at the $\Gamma$ point in five-layer phosphorene is decreased to a value of 0.82 eV. Similar arguments show that, in this case, a relatively small critical strain will be needed to generate a large BCD peak at the Fermi level. Our calculations support this conjecture and reveal the value of the critical tensile strain to be 6.5\% [Fig.~\ref{fig:bcd2} (d)], which is almost 1\% less compared to the three-layer. The sign change of the BCD peak is also found in five-layer phosphorene systems; however, the peak values, in this case, are substantially larger. It is evident from Fig.~\ref{fig:bcd2} (d) that a large BCD peak (-15.3 \AA{}) is observed at the Fermi level with a negative sign for the strain value of 7.1\%. This BCD peak gradually changes its sign to a positive value of 10.6 \AA{} for 8\% strain. Moreover, multiple Dirac-like closely separated bands emerge around the strain of $\approx 8.9$\%, and multiple adjacent peaks are observed at the Fermi level. Therefore, we discover that increasing the number of phosphorene layers can reduce the critical strain value for the generation of a strong BCD peak at the Fermi level. Furthermore, the multilayer phosphorene systems exhibit additional large BCD peaks within the experimentally accessible energy range, which were absent in the monolayer case. The values of the BCD obtained in these systems are significant. For a comparison, we note that the obtained values of BCD in strained monolayer MoS$_2$~\cite{son2019strain} and $T_d$ phase of monolayer WTe$_2$~\cite{you2018berry}
are of the order of $ \approx \pm 10^{-2}$ \AA{} and $\approx \pm 10^{-1}$ \AA{}, respectively. Remarkably, the sign of BCD can also be tuned using an appropriately chosen strain.

\section{Conclusions and outlook}

In a nutshell, we have presented strain engineering as a promising approach for achieving a strong BCD in layered phosphorene systems. First, we have proposed a staggered onsite potential to break the inherent inversion symmetry of phosphorene -- this will allow the system to possess a non-zero Berry curvature even under time-reversal symmetric conditions. In practice, phosphorene can experience such an inversion symmetry breaking potential in the presence of an appropriate substrate, for example, monochalcogenides~\cite{hu2019recent}. Alternatively, the combined effect of external and lattice commensurate electric fields can also induce such an effect. The staggered onsite potential not only breaks the inversion symmetry but also reduces the crystal symmetry to a single mirror line. Consequently, the criteria for obtaining a BCD in 2D systems are fulfilled in phosphorene. Nevertheless, the wide band gap in monolayer ($\approx 1.52$ eV) restricts the emergence of any BCD peak near the Fermi level. We next proposed to apply an external tensile strain to overcome this drawback. The strain essentially reduces the band gap, which causes a systematic enhancement of the BCD. Finally, near the strain value of 11.6\%, two BCD peaks associated with the top of the valence band and bottom of the conduction band merge, resulting in a significantly strong BCD response at the Fermi level. We have increased the strain further, which gives rise to the most intense BCD peak of 5.6 \AA{} at 11.8\% strain. It is worth mentioning that the inversion symmetric phosphorene undergoes a strain-driven topological phase transition about a critical gap closing point at the same 11.8\% strain. However, in the broken inversion symmetric phase, the BCD peak gradually changes its sign around this point. Although phosphorene can sustain tensile strain up to substantially large values, increasing the layer number can reduce the strain required for a strong BCD peak at the Fermi level. The underlying reason behind the reduction of critical strain values in layered phosphorene is the associated decreasing trend of the band gap. For instance, the band gap of three-layer (0.95 eV) and five-layer (0.82 eV) phosphorene is almost half of that of the monolayer. Consequently, the critical strain value for the three-layer and five-layer cases reduce to 7.4\% and 6.5\%, respectively. Therefore, we have discovered that the inversion symmetry broken phosphorene systems can exhibit strong BCD peaks at the Fermi level that can also change their sign after strain engineering. Moreover, the required strain value can be reduced to less than $10\%$ by increasing the number of layers, say, three- or five-layered phosphorene. By proposing this new, tunable materials platform, we are hopeful that our work will pave the way to realize the nonlinear Hall effect in the 2D phosphorene systems under experimentally attainable conditions.

The nonlinear Hall response mentioned above has several practical applications that can prove the fascinating properties of emergent quantum materials. For example, in spintronics, the spin and charge conversion efficiency is primarily determined by Berry curvature induced spin Hall conductivity. Different quantum numbers, such as spin and orbital, can cause Berry curvature and BCD in a system, namely spin-sourced BCD and orbital-sourced BCD. These distinct sources manifest a combined effect of spintronics and optoelectronic responses in a single system~\cite{lesne2023designing}. In other words, orbital-sourced Berry curvature contributes to the photogalvanic currents that can trigger spin Hall voltage in the presence of spin-sourced Berry curvature. Experimental evidence of the nonlinear Hall effect opens the avenue for realizing the higher-order spin-to-charge conversion beyond the linear-response theories~\cite{kozuka2021observation}. Furthermore, the local spin-charge interconversion can also be achieved by fabricating suitable van der Waals heterostructures with noticeable spin-orbit coupling effects~\cite{powalla2022berry}. The current-induced spin polarizations can further be tuned by gate voltages that persist even up to room temperature. It is interesting to note that spin polarization has a prominent effect on the Kerr signals, which is measurable by angle-dependent study of the Kerr response, i.e., Kerr rotation microscopy. This appearance of the Kerr signal in experiments is essentially a fingerprint of non-zero BCD in the sample. Moreover, spin-momentum locking, band tilting and other symmetry-related effects such as Fermi surface warping give rise to spin-polarized nonlinear Nernst effects~\cite{zeng2022band}. Further, the anomalous nonlinear Hall effect is also well equipped to detect the orientation of the N \`eel vector and its electrical control by spin-orbit torques in the case of antiferromagnetic spintronics. The orientation of N\`eel vector in antiferromagnets strongly depends on the BCD of these antiferromagnetic systems that invariably results in a prominent nonlinear anomalous Hall effect~\cite{shao2020nonlinear, wang2021intrinsic}. In a recent experiment, a nonlinear Hall response has been explored after interfacing black phosphorus with even-layered parity-time symmetric antiferromagnet MnBi$_2$Te$_4$~\cite{gao2023quantum}. The nonlinear Hall response is non-dissipative and flipped in its direction when the spins of the MnBi$_2$Te$_4$ are flipped. In this work, we have discovered that strained engineered phosphorene exhibits a pronounced BCD compared to the pristine one, and it can also be externally switchable. This enhanced BCD and the associated nonlinear Hall effect of strained phosphorene and interface formation with antiferromagnetic or high spin-orbit coupling systems will provide an efficient platform for spintronics applications. Therefore, a synergy between nonlinear transport and spintronics leads to various exciting properties that may be useful in applications.

\section*{Acknowledgements}

A.B. sincerely acknowledges the financial support from Indian Institute of Science IoE postdoctoral fellowship. N.B.J. acknowledges Prime Minister’s Research Fellowship for support. A.N. acknowledges support from the start-up grant (SG/MHRD-19-0001) of the Indian Institute of Science.

\bibliographystyle{apsrev}
\bibliography{ref}

\end{document}